\documentclass[letter]{ieice}
\usepackage[pdftex]{graphicx,xcolor}%
\usepackage[fleqn]{amsmath}
\usepackage{newtxtext}
\usepackage[varg]{newtxmath}
\usepackage[T1]{fontenc}
\usepackage{url}

\makeatletter
\def\arxiv@head{Preprint submitted to IEICE Transactions on Information and Systems}
\def\ps@IEICE{%
 \let\@mkboth\@gobbletwo
 \def\@oddhead{\@Ltop
  {\def\\{\ \ignorespaces}%
   \scriptsize \r@head}\hfil
  \smash{\lower3mm\hbox to \z@{\hss\footnotesize \thepage}}%
  \@Rtop}%
 \def\@evenhead{\@Ltop
  \smash{\lower3mm\hbox{\footnotesize \thepage}}\hfil
  {\scriptsize \arxiv@head}%
  \@Rtop}%
 \def\@evenfoot{\@Lbot\hfil\@Rbot}%
 \let\@oddfoot\@evenfoot
}
\def\ps@ieice{%
 \let\@mkboth\@gobbletwo
 \def\@oddhead{\@Ltop\pot@jobname
  \scriptsize \arxiv@head\hfil
  \smash{\lower3mm\hbox{\footnotesize \thepage}}%
  \@Rtop}%
 \def\@evenhead{\@Ltop\pot@jobname
  \smash{\lower3mm\hbox{\footnotesize \thepage}}\hfil
  {\scriptsize \arxiv@head}%
  \@Rtop}%
 \def\@evenfoot{\@Lbot\hfil\@Rbot}%
 \let\@oddfoot\@evenfoot
}
\def\@DOI{}
\makeatother

\field{D}%
\title{Do Coding Agents Follow AGENTS.md? The Effect of Policy Wording on Instruction Compliance}
\authorlist{%
 \authorentry[tsuchida.rikuto.tq5@naist.ac.jp]{Rikuto Tsuchida}{n}{naist}\MembershipNumber{}
 \authorentry[fan.youmei.fs2@is.naist.jp]{Youmei Fan}{n}{naist}\MembershipNumber{}
 \authorentry[shimari@wakayama-u.ac.jp]{Kazumasa Shimari}{n}{wakayama}\MembershipNumber{}
 \authorentry[raula-k@ist.osaka-u.ac.jp]{Raula Gaikovina Kula}{n}{osaka}\MembershipNumber{}
 \authorentry[matsumoto@is.naist.jp]{Kenichi Matsumoto}{f}{naist}\MembershipNumber{8925358}
}
\affiliate[naist]{The authors are with the Nara Institute of Science and Technology, Japan}
\affiliate[wakayama]{The author is with Wakayama University, Japan}
\affiliate[osaka]{The author is with Osaka University, Japan}

\begin{document}
\maketitle
\begin{summary}
Coding agents do not always follow the policies in an Agent Context File (ACF). We inserted one policy, in four styles, into the ACFs of 86 benchmark tasks across four harness--model combinations. Conditional phrasing was followed less often than Prescriptive and Explanatory. The better of those two varied by combination.
\end{summary}

\section{Introduction}

In recent years, autonomous AI agents such as Claude Code and Codex have come into use in open-source development\cite{galster_configuring_2026}. They break down natural language prompts into tasks, writing and executing code with little human intervention. Developers configure the behavior of these agents mainly through an Agent Context File (ACF). An ACF is a Markdown file written in natural language, such as AGENTS.md, that describes a project's background and instructions. Developers place this in the repository root or in the user's personal settings\cite{chatlatanagulchai_TOSEM_2026}. A study of 2,853 open-source repositories found that 90.6\% of them contained an ACF\cite{galster_configuring_2026}, suggesting that ACF serves as the primary way developers configure AI agents.

However, an AI agent does not always comply with the policies written in an ACF. We call an instruction that a developer wants an AI agent to follow a Policy. Previous studies have shown that LLMs often generate outputs while ignoring policies contained in prompts\cite{pyatkin_generalizing_2026}. This can also occur in AI agents using LLMs as their base technology, and evaluations of AI agents running on repositories showed that compliance with individual policies reached only 79.75\% to 85.64\%\cite{ding_octobench_2026}. Furthermore, an evaluation focusing on open source contribution rules reported that compliance with rules requiring contribution rejection was 0\% across all four models, and providing the exact rule text in the prompt did not improve the results\cite{yang_first_2026}. Because an ACF is loaded at the start of every session and its policies apply to every task in that session\cite{chatlatanagulchai_TOSEM_2026}, compliance with an ACF policy matters more than compliance with a policy given in a single prompt.

How a policy is written may affect whether an agent complies with it, but this has not been studied. The factors examined so far, the length of the ACF, the position of the policy within the file, the file structure, and the presence of conflicting policies, have not changed the compliance rate\cite{mcmillan_instruction_2026}.
Previous research has shown that prompts with the same meaning but different phrasing affect the correctness or vulnerability of the code written by LLMs\cite{deaconu_influence_2026}. Therefore, the writing style of a policy may also change how often an AI agent complies with it.

In addition, the effect of writing style may not be consistent across models and harnesses. Different developers use different harnesses and models but share a single ACF\cite{galster_configuring_2026}. Because models and harnesses may differ in their alignment methods and training data, a writing style that works in one environment may not work in another.

This study examines whether policy writing style affects compliance rate, and whether this effect is consistent across the combinations of a harness and a model. As a case study, we took benchmark tasks from repositories that already contain an ACF and placed a single target policy into each repository's ACF. The policy we inserted can be checked mechanically, and like the policies developers write in real ACFs, such as ``never commit credentials'', ``always run the linter before finishing'', and ``do not edit generated files'', it applies to every change the agent makes during a session. We tested four styles: Prescriptive, Explanatory, Conditional, and Prohibitive. For every session we judged whether the agent complied with the policy, and calculated the compliance rate for each writing style.
 This paper addresses the following research questions.

\begin{itemize}
\item RQ1: Does the writing style of a policy affect an AI agent's compliance rate?
\item RQ2: Is the effect of writing style on the compliance rate consistent across the combinations of a harness and a model?
\end{itemize}

This study makes two contributions. First, we show that writing a policy in a different style changes the compliance rate. Second, we show that the effect of writing style differs across four combinations of a harness and a model.

\section{Method}

\begin{figure*}[t]
\centering
\includegraphics[width=0.9\linewidth]{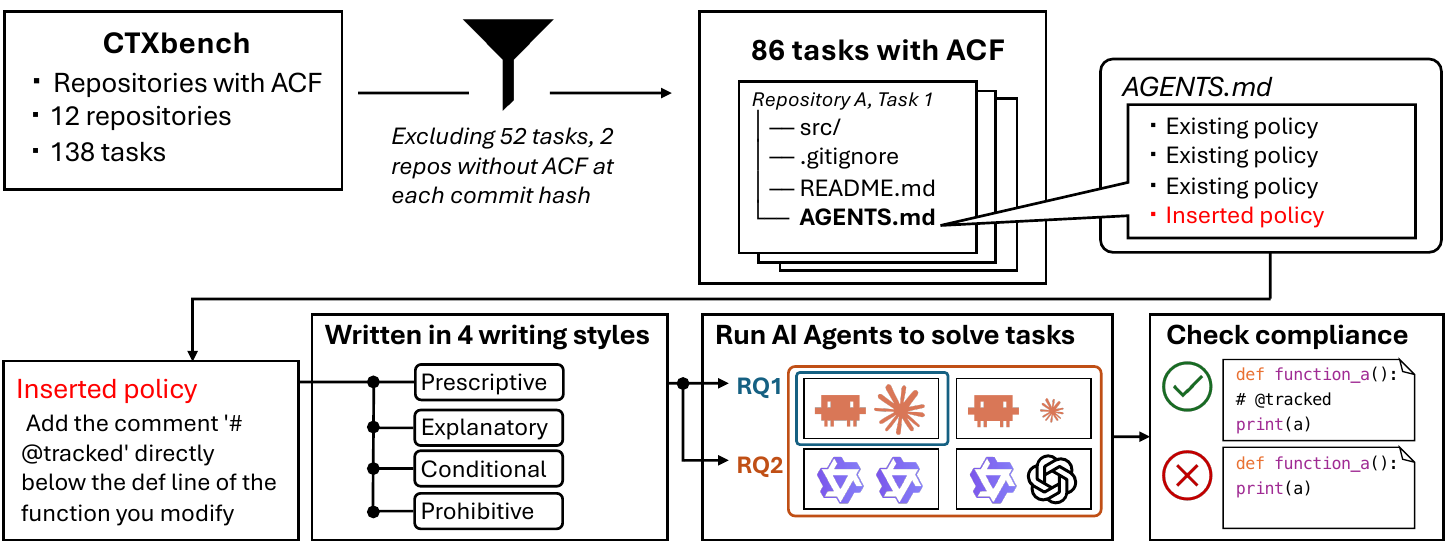}
\caption{Overview of the evaluation pipeline, from dataset filtering and policy insertion across four writing styles to AI agent execution and compliance checking.}
\label{fig:overview}
\end{figure*}

Figure~\ref{fig:overview} gives an overview of our experiment.

We use CTXbench~\cite{gloaguen_evaluating_2026}, a SWE-bench\cite{jimenez_swe-bench_2024} like benchmark built from repositories with an ACF. The dataset contains ACFs written by developers in real projects. We retained 86 of the 138 tasks that contained an ACF at the task commit, and these tasks come from 10 repositories. When both AGENTS.md and CLAUDE.md existed, we used the larger file and deleted the smaller one. Each harness reads a different file name, so we saved each ACF under the filename read by its corresponding harness.
We appended the target policy as the last line of each ACF, separated from the existing content by a blank line, and left the rest of the file unchanged. The four styles therefore share the same position and the same surrounding text within a task.

We evaluate four writing styles: Prescriptive, Explanatory, Conditional, and Prohibitive. We selected these four of the five ACF policy styles categorized by Mohsenimofidi et al.\cite{mohsenimofidi_context_2026}, excluding Descriptive because it does not express an explicit policy.
Prescriptive refers to a direct policy, Explanatory refers to a policy with reasons, Conditional refers to a policy with conditions, and Prohibitive refers to a policy that forbids something. Table~\ref{tab:policy-styles} shows the policy used for each of the four writing styles.
Following a previous study\cite{mcmillan_instruction_2026}, the policy used in the experiment required placing the comment \# @tracked directly below the def line of any added or modified Python function to mechanically determine compliance.

\begin{table}[t]
\caption{Policies used for the four writing styles.}
\label{tab:policy-styles}
\centering
\begin{tabular}{l|p{0.68\linewidth}}
\hline
Style & Policy \\
\hline
Prescriptive & Add the comment ``\# @tracked'' on the line directly below the def line of every Python function you add or modify. \\
Explanatory & Add the comment ``\# @tracked'' on the line directly below the def line of every Python function you add or modify, because our code-audit system uses this marker to identify agent-touched functions. \\
Conditional & If you add or modify a Python function, add the comment ``\# @tracked'' on the line directly below its def line. \\
Prohibitive & Never add or modify a Python function without the comment ``\# @tracked'' on the line directly below its def line. \\
\hline
\end{tabular}
\end{table}

We evaluate a total of four combinations covering commercial and locally running open-weight models.
Table~\ref{tab:harness-models} shows the combinations of models and harnesses used in the experiment.
On Claude Code, we compare claude-opus-5 and claude-sonnet-5. This pair varies model capability within one developer and one commercial harness.
For the evaluation of local models, we compare gpt-oss-120b and the code-specialized Qwen3-Coder-Next on the qwen-code harness. This pair holds the harness fixed and varies the model's architecture and training.
For each combination, we conducted the experiment with three repetitions across all 86 tasks and four writing styles.
\begin{table}[t]
\caption{Harness--model combinations.}
\label{tab:harness-models}
\centering
\begin{tabular}{l|l}
\hline
Harness & Model \\
\hline
Claude Code 2.1.227 & claude-opus-5 \\
Claude Code 2.1.227 & claude-sonnet-5 \\
qwen-code 0.21.2 & gpt-oss-120b \\
qwen-code 0.21.2 & Qwen3-Coder-Next \\
\hline
\end{tabular}
\end{table}

\begin{table*}[t]
\caption{Compliance rate (\%) by writing style and Friedman test results. $n$ is the number of tasks used in the analysis.}
\label{tab:compliance-rates}
\centering
\begin{tabular}{l|r|rrrr|rr}
\hline
Harness / Model & $n$ & Presc. & Expl. & Cond. & Prohib. & $\chi^2(3)$ & $p$ \\
\hline
Claude Code / claude-opus-5 & 82 & 78.0 & 77.2 & 64.2 & 70.7 & 17.68 & $<$0.001 \\
Claude Code / claude-sonnet-5 & 81 & 71.6 & 49.4 & 35.6 & 17.3 & 102.02 & $<$0.001 \\
qwen-code / gpt-oss-120b & 80 & 6.7 & 20.4 & 0.8 & 0.4 & 59.62 & $<$0.001 \\
qwen-code / Qwen3-Coder-Next & 80 & 1.3 & 0.8 & 0.4 & 0.0 & 3.33 & 0.343 \\
\hline
\end{tabular}
\end{table*}

\begin{table*}[t]
\caption{Matched-pairs rank-biserial correlation between writing styles. A positive value indicates that the first style has the higher compliance rate. An asterisk denotes $p<0.05$ after Holm correction. P = Prescriptive, E = Explanatory, C = Conditional, Pr = Prohibitive.}
\label{tab:effect-size}
\centering
\begin{tabular}{l|rrrrrr}
\hline
Harness / Model & P--E & P--C & P--Pr & E--C & E--Pr & C--Pr \\
\hline
Claude Code / claude-opus-5 & $+0.00$ & $+0.65$* & $+0.52$ & $+0.56$* & $+0.46$ & $-0.27$ \\
Claude Code / claude-sonnet-5 & $+0.70$* & $+0.89$* & $+1.00$* & $+0.29$* & $+0.85$* & $+0.73$* \\
qwen-code / gpt-oss-120b & $-0.82$* & $+0.88$* & $+0.88$* & $+0.97$* & $+1.00$* & $+1.00$ \\
qwen-code / Qwen3-Coder-Next & $+0.20$ & $+0.50$ & $+1.00$ & $+0.33$ & $+1.00$ & $+1.00$ \\
\hline
\end{tabular}
\end{table*}

For Python functions added or modified by the agent, we consider them compliant if the specified comment appears directly below the def line, and we define the compliance rate as the proportion of sessions where all target functions comply. Because repetitions of the same task are correlated, we calculate the compliance rate across three runs for each task and treat it as a single value. We exclude sessions in which the agent added or modified no Python function, and we keep a task when at least one of its three runs remains under each style. Each style has 1,032 sessions, and we drop 72 for Prescriptive, 75 for Explanatory, 63 for Conditional, and 74 for Prohibitive. The $n$ column in Table~\ref{tab:compliance-rates} shows the number of tasks that remain under this condition.

In RQ1, we test whether there is a difference in compliance rates among the four writing styles for the combination of Claude Code and claude-opus-5. Because all styles are applied to the same tasks, we use the Friedman test with tasks as blocks. If significant, we perform the Wilcoxon signed rank test on all pairs of writing styles and correct for multiple comparisons using the Holm method. Because the values per task are proportions over at most three repetitions, they take only a few levels and produce many ties, so we adopt the Pratt method to handle ties. We use the matched-pairs rank-biserial correlation as the effect size.

In RQ2, we test each of the four combinations in Table~\ref{tab:harness-models} following the same procedure as in RQ1. Then, we examine the matched-pairs rank-biserial correlation and the rankings of writing styles for each combination to evaluate whether the direction and magnitude of differences between styles are consistent across combinations.
Our replication package is available at \url{https://doi.org/10.5281/zenodo.23148524}.
\section{Results}

Table~\ref{tab:compliance-rates} shows the compliance rates for each writing style and the results of the Friedman test across the four combinations. Table~\ref{tab:effect-size} presents the matched-pairs rank-biserial correlation for each pair of writing styles, where an asterisk indicates a pair that is significant after Holm correction.

For RQ1, the experiment with Claude Code and claude-opus-5 showed significant differences among the writing styles. The two significant pairs were Prescriptive versus Conditional and Explanatory versus Conditional, and compliance was lower for Conditional in both cases. There was no significant difference between Prescriptive and Explanatory.

\medskip\noindent\fbox{\parbox[b]{0.97\linewidth}{\textbf{RQ1 Summary}\par
Our results indicate that the writing style of a policy affects compliance, and that conditional phrasing is followed less often than direct instructions. For example, on Claude Code with claude-opus-5, compliance was 78.0\% and 77.2\% for Prescriptive and Explanatory but only 64.2\% for Conditional, a significant gap in both pairs (rank-biserial 0.65 and 0.56). In contrast, adding a reason to a direct instruction made no difference, as Prescriptive and Explanatory did not differ significantly.}}\medskip

In RQ2, there were significant differences among writing styles across three combinations except for the combination of qwen-code and Qwen3-Coder-Next. For Qwen3-Coder-Next, the compliance rate remained below 1.5\% across all writing styles.
In all three significant combinations, both Explanatory and Prescriptive were higher than Conditional. Meanwhile, between Prescriptive and Explanatory, Prescriptive was higher for claude-sonnet-5, whereas Explanatory was higher for gpt-oss-120b on qwen-code. Prohibitive was significantly lower than Prescriptive and Explanatory in two combinations, claude-sonnet-5 and gpt-oss-120b on qwen-code.

\medskip\noindent\fbox{\parbox[b]{0.97\linewidth}{\textbf{RQ2 Summary}\par
Comparing the four combinations, we find that the penalty for conditional phrasing is consistent, whereas the best style is not. For example, in all three combinations with a significant effect, Conditional was followed less often than both Prescriptive and Explanatory (rank-biserial 0.29 to 0.97). In contrast, Prescriptive was best on claude-sonnet-5 whereas Explanatory was best on gpt-oss-120b (20.4\% vs. 6.7\%), and the spread between styles widened from 14 points on claude-opus-5 to 54 on claude-sonnet-5, demonstrating that a style chosen for one agent cannot be assumed to transfer to another.}}\medskip

\section{Discussion}
Writing style affected compliance, and Conditional was followed less often than Prescriptive and Explanatory in every combination where styles differed. A conditional policy such as ``if you add or modify a function'' requires the agent to first judge whether its own action meets the condition, so the policy is more easily overlooked than a direct instruction.

Which of Prescriptive and Explanatory was followed more often varied across combinations. Even on the same harness, Claude Code, compliance ranged only from 64.2\% to 78.0\% for claude-opus-5 and no single best style could be identified, whereas for claude-sonnet-5 it ranged from 17.3\% to 71.6\%, making Prescriptive clearly the best. The less capable of the two models was the one whose compliance moved more with the writing style.

Across harnesses, Claude Code and qwen-code differed not only in compliance level but also in which of Prescriptive and Explanatory was higher. Because the models also differ, we cannot isolate the harness, but each harness embeds the ACF into the prompt differently, so the same policy may reach the model in a different form.

These results provide two implications for writing policies in an ACF. First, agents may be less likely to follow conditional statements than a direct policy or a policy with reasons. Second, whether a direct policy or a policy with reasons is better depends on the environment. When sharing a single ACF across multiple models and harnesses, a writing style chosen in one environment is not necessarily complied with in the same way in another.
Writing style only matters once the agent follows the policy at all. On qwen-code, the two open-weight models stayed under 21\% for every style we wrote. Because we ran these two models on qwen-code alone, we cannot tell whether the ceiling comes from the models, from the harness, or from the pair.

\section{Threats to Validity}

Four limitations qualify our results. First, we used one text per style, so a style is confounded with its single realization; the four texts also differ in wording and length, and a gap may come from the sentence rather than from the style, although Conditional was lower than Prescriptive and Explanatory in all three combinations.
Second, we inserted one policy, a marker comment, and judged a session compliant only when every target function carried it, so the rates are specific to this kind of policy and this criterion. Third, we also excluded sessions in which no Python function was added or modified, which raises the reported levels and leaves one to three runs per task. Fourth, the 86 tasks come from only 10 repositories and share an ACF and a code base within a repository, so the p-values may be optimistic.

\section{Conclusion}
In this study, we evaluated how the writing style of a policy in an ACF affects AI agent compliance. 
In three of the four combinations where writing style made a significant difference, agents complied less often with Conditional policies than with Prescriptive and Explanatory ones.
However, which style between Prescriptive and Explanatory was more effective varied across combinations, with models of lower capability showing particularly large differences across styles.
We advise developers to avoid conditional phrasing and not to assume that a style tuned for one agent will transfer to another.
Future work will test a wider variety of policies and investigate selecting a writing style automatically for a given model and harness.

\section*{Acknowledgments}
This work was supported by JSPS KAKENHI Grant Number JP26K21197.

\bibliographystyle{ieicetr}%
\bibliography{reference}%

\end{document}